\documentclass[12pt]{article}
\usepackage{epsfig}
\usepackage{psfig}
\begin{document}
\begin{center}
{\Large{
Some Theoretical Aspects of Quantum Mechanical Equations  in Rindler Space}}

\bigskip
{\large{
Soma Mitra, Sanchari De and Somenath Chakrabarty$^\dagger$

\medskip
Department of Physics, Visva-Bharati, Santiniketan-731235,
India\\
$^\dagger$somenath.chakrabarty@visva-bharati.ac.in}}


\end{center}

\bigskip
\begin{center}
\bf{Abstract}
\end{center}
In this article we have investigated some of the theoretical aspects of the solutions of quantum mechanical equations 
in Rindler space. We have developed the formalism for exact analytical solutions for Schr$\ddot{\rm{o}}$dinger equation and Klein-Gordon
equation. Along with the approximate form of solutions for these two quantum mechanical equations. We have discussed the physical
significance of our findings. The
Hamiltonian operator in Rindler space is found to be non-Hermitian in
nature. But the energy
eigen values or the energy eigen spectra are observed to be real. We
have noticd that the
sole reason behind
such real behavior is the PT symmetric form of
the Hamiltonian operator.

\section{introduction}
From the knowledge of our literature survey of the articles on
general relativity and related topics, we have
noticed that
the principle of equivalence plays a vital role in the studies of
various aspects of classical and quantum physics in a uniformly
accelerated frame or in  Rindler space \cite{R1,R2,R3,R4} in which
the backgound gravitational field is uniform. According to this
principle, a frame undergoing uniform accelerated motion in absence of
gravity is equivalent to a frame at rest in presence of a constant
gravitational field. However, the strength of gravitational field can not
be constant throughout the whole space. Within a limited region it is uniform.
Hence the constant acceleration of the frame is also called the local acceleration. Assuming that the uniform
acceleration is along $x$-direction, the metric in the Rindler space is then given by
\begin{equation}
g^{\mu \nu} \equiv \left( \left ( 1 + \frac{\alpha x}{c^2} \right )cdt , dx , 0 , 0 \right)
\end{equation}
where $\alpha$ is the uniform acceleration. In the {\bf{A1}} part of the
Appendix, following \cite{R5,R6,R7,R8}, 
we have given a brief derivation of various physical quantities.
As a comparison, in the Minkowski space the metric is given by
\begin{equation}
g^{\mu \nu} \equiv \left( cdt , dx , 0 , 0 \right)
\end{equation}
Hence the line element in the $1+1$-dimensional Rindler space is given by
\begin{equation}
ds^2 = \left( 1 + \frac{\alpha x }{c^2} \right)^2 (cdt)^2 - dx^2
\end{equation}
and
\begin{equation}
ds^2 = (cdt)^2 -dx^2
\end{equation}
in the $1+1$-dimensional Minikowski space.
To obtain the Lagrangian and Hamiltonian of the system, we proceed
exactly the same manner as have been done in special theory of
relativity (see {\bf{A1}} part of the Appendix). The
Rindler space is essentially associated with a uniformly accelerated
frame, otherwise
it is like flat Minikwoski space. Then following Landau and Lifshitz \cite{R1}, we have
the action integral
\begin{equation}
S = - \alpha _0 \int_{1}^{2}ds = - \alpha _0 \int_{1}^{2} cdt \left[
\left( 1 + \frac{\alpha x}{c^2} \right)^2 - \frac{v^2}{c^2}
\right]^{\frac{1}{2}}
\end{equation}
where $ v = \frac{dx}{dt} $, the three velocity of the particle.
We sustitute $\alpha_0 = m_0 c $, as has been done in the case of special
theory of relativity, where $m_0$ is the particle rest mass. Then we have 
\begin{equation}
S = m_0 c^2 \int_{1}^{2} dt \left[ \left( 1 + \frac{\alpha x}{c^2}
\right)^2 - \frac{v^2}{c^2} \right] = \int_{1}^{2} Ldt
\end{equation}
From the above equation, the classical Lagrangian of the particle is
given by
\begin{equation}
L = - m_0 c^2 \left[ \left( 1 + \frac{\alpha x}{c^2}\right)^2 -
\frac{v^2}{c^2} \right]^{\frac{1}{2}}
\end{equation}
Now from the conventional definitions,
we have the particle three momentum 
\begin{equation}
p = \frac{dL}{dv} = \frac{m_0 \vec{v}}{\left[ \left( 1 +\frac{\alpha
x}{c^2} \right)^2 -
\frac{v^2}{c^2} \right]^{\frac{1}{2}}}
\end{equation}
and the corresponding Hamiltonian 
\begin{equation}
H = pv - L = \left( 1 + \frac{\alpha x }{c^2} \right) \left( p^2 c^2 +
m_0^2 c^4 \right)^{\frac{1}{2}}
\end{equation}
It is quite evident from eqns.(6)-(9) that for the inertial frame with
$ \alpha = 0 $, we get back the results of special theory of relativity.
In the case of classical mechanics, $ x$, $p$ and  $H$  are dynamical
variables, whereas in the quantum mechanical picture, $ x$, $p$ and $H $
are operators. In the later case $ x$ and  $p $ are also canonical conjugate of each other, i.e., $
[ x , p ] = i \hbar$. Hence it is quite obvious that the Hamiltonian operator
represented by eqn.(9) is non-Hermitian. However from our subsequent
analysis and discussion we will show that the eigen values or the eigen
spectra are real in nature. This is found to be solely because of the $PT$ symmetric
nature of the Hamiltonian operator \cite{PT}. 
Under $ P $ and $ T $ operations we have the following relations from $PT$-symmetric 
quantum mechanics: $ p x p^{-1} = -x$, $T x T^{-1} = x$, $P p P {^-1} = - p$,
$T p T ^{-1} = -p$, $P \alpha P^{-1} = - \alpha$, $T \alpha T^{-1} = \alpha$  and
$T i T ^{-1} = - i$. The last relation is essential for the  preservation
of canonical quantization relation under $PT$ operation, i.e., for
the validity of $ PT [x , p] (PT)^{-1} = i \hbar $.
Hence it is quite obvious to verify that for the Hamiltonian, given by eqn.(9), $ PT H (PT)^{-1} = H $, i.e., the
Hamiltonian operator is $ PT $ invariant. We shall show in our subsequent
discussion that since eigen functions $\Psi$ are the functions of the product $
\alpha x $, which is $ PT $ symmetric, therefore $ PT \Psi (u) = \Psi(u) $,
where $ u $ is a function of the product of $ \alpha x $. 

To study some of the quantum aspects in Rindler space, we start with
the Hamiltonian given
by eqn.(9). In this article along with the exact relativistic solution of Klein-Gordon equation,
we shall also make non-relativistic
approximation and solve for Schr$\ddot{\rm{o}}$dinger equation in an exact manner. We have further 
solved the relativistic and non-relativistic
quantum mechanical equations with some approximations. We have discussed the physical significance of the solutions.
To the best of our knowledge such studies have not been done before,
except some preliminary studies by our group \cite{SSS1,SSS2}. For the sake of completeness we
have also discussed two of our already reported works. 

We have organized the article in the following manner. In the next
section, we shall make a non-relativistic approximation for the
Hamiltonian and developed a formalism for the exact solution of the
schr$\ddot{\rm{o}}$dinger equation in Rindler space
and given the physical interpretation of our results. In section $ 3$, we have solved the
non-relativistic equation with some approximation and shown the
analogy between our result with the cold field emission of electrons
from a metal surface under the action of a strong electric field
\cite{FN} (see also \cite{AS}). In
section $ 4 $, we have developed the formalism for another kind of
approximate solution for the
schr\"{o}dinger equation. In section $ 5 $, we have presented an exact
solution for the relativistic form of quantum mechanical equation. In
section $ 6 $, we have solved the relativistic Klein-Gordon equation.
The later formalism is also exact in nature. Finally we have given
the conclusion of our work.
\section{ Schr\"{o}dinger Equation in Rindler Space (An Exact Solution)}
In the non-relativistic approximation the Hamiltonian is given by
\begin{equation}
H \approx \left( 1 + \frac{\alpha x}{c^2}\right)\left(m_0c^2 + \frac{p^2}{2m_0}\right)
\end{equation}
Then the Schr$\ddot{\rm{o}}$dinger equation, $ H\Psi = E\Psi $, may be expressed
as
\begin{equation}
\left( 1 + \frac{\alpha x }{c^2} \right)\left(m_0c^2 + \frac{p^2}{2m_0}
\right)\Psi(x , y , z) = E \Psi(x , y, z)
\end{equation}
Since there is no $ y $ or $ z $ dependent terms in the Hamiltonian,
the separable form of the wave function can be written in the form
\begin{equation}
\Psi(x , y, z) = N \exp\left (-\frac{ip_y y }{\hbar}\right )\exp\left (-\frac{ip_z z
}{\hbar}\right )X(x)
\end{equation}
where $ N $ is the normalization constant. Substituting $ \Psi(x , y ,
z) $ in eqn.(11), we have
\begin{equation}
\left( 1 + \frac{\alpha x }{c^2}\right) \left( -
\frac{\hbar^2}{2m_0}\frac{d^2}{dx^2} + E_\perp \right) X(x) = EX(x)
\end{equation}
where 
\begin{equation}
E_\perp = \frac{p_y ^2 + p_z ^2}{2m_0} +m_0c^2
\end{equation}
the transverse part of particle energy. To solve the above
differential equation (eqn.(13)), let us make a coordinate
transformation, given by
\begin{equation}
u = 1 + \frac{\alpha x}{c^2} 
\end{equation}
Then the Schr\"{o}dinger equation (eqn.(13)) reduces to the following form
\begin{equation}
\frac{d^2 X}{du^2} + \frac{2m_0 c^4}{\hbar ^2 \alpha ^2} \frac{E}{u}X
-\frac{2m_0 c^4}{\hbar ^2 \alpha ^2} E_\perp X = 0
\end{equation}
which may also be written in the form 
\begin{equation}
\frac{d^2 X}{dw^2} +\left( -\frac{1}{4} + \frac{\gamma}{w} \right) X(w) = 0
\end{equation}
where $ w = b^{1/2}u , ~\gamma = ab^{1/2},~ ~a =
\frac{2m_0 c^4}{\hbar ^2 \alpha^2}E ~ ~{\rm{and}}~ ~b= \frac{8m_0 c^4}{\hbar ^2
\alpha ^2}E_\perp $.
This is the differential equation satisfied by an one dimensional
hydrogen atom, or in other words the eigen value problem in Rindler
space is equivalent to one dimensional quantum mechanical
hydrogen atom problem \cite{ODH1,ODH2}. 
On comparing the above differential equation (eqn.(17))
with that satisfied by the Whittaker function $ W_{k , \mu} (x) , $
given by \cite{MB}
\begin{equation}
\frac{d^2}{dx^2}M_{k , \mu}(x) + \left( -\frac{1}{4} +\frac{k}{x}
+\frac{\frac{1}{4} - \mu ^2}{x^2} \right)M_{k, \mu}(x) = 0
\end{equation}
we have $ X(x)\equiv M_{k,\mu}(x)$ and for this particular problem $
x=w , \mu = \frac{1}{2} 
~{\rm{and}}~ k = \gamma$. Then
\begin{equation}
X(w ) = M_{\gamma , \frac{1}{2}}(w)= \exp \left( -\frac{w}{2} \right) w
M(1-\gamma , 2 , w)
\end{equation}
where $M(a , c , x)= {_1F_1 (a ; c; x)}$, is the confluent Hyper-geometric
function \cite{R11}. Now the hyper-geometric function $ M(1-\gamma , 2 ,w ) $ will
be a polynomial and becomes zero for $ w \longrightarrow \infty $, if
the parameter $ 1-\gamma $ is zero or a negative integer, i.e., $ \gamma = n
$, for  $ n = 1 , 2 , 3 ,......,$, the positive integers \cite{MB}. This is 
the physical condition for the bounded nature of the wave function along
positive $ x-$direction. 
Under such restricted situation, the solution
can also be expressed in terms of Associated Laguerre function
\cite{ODH1,ODH2}. This alternative form of  wave function is then given by
\begin{equation}
X(w)=\exp\left (-\frac{w}{2}\right )w L_{\gamma-1}^1(w)
\end{equation}
The parameter $\gamma$ is again have to be non-zero positive integer. It can very easily be verified that
the structure of eqn.(17) is exactly identical with the equation for an 
one-dimensional hydrogen atom \cite{ODH1,ODH2}.
Then it is just a matter of simple algebra to show that under such restricted condition we get
quantized form  of energy of the particle in an
uniformly accelerated frame. Using the expressions for $a$, $b$ and 
$E_\perp$, it is straight forward to
show that the quantized form of energy of the particle is given by 
\begin{equation}
E_n=n\hbar\frac{\alpha}{c} =n\hbar \omega ~~{\rm(say)}
\end{equation}
with 
\begin{equation}
\omega=\frac{\alpha}{c}
\end{equation}
where we have dropped $2^{1/2}$ term for aesthetic ground and neglected 
$p_y$ and $p_z$ for a purely one-dimensional condition. Then accordingly $E_\perp=m_0c^2$, the rest mass energy of
the particle. Obviously it is quite
surprising result. The differential equation satisfied 
by the particle is exactly identical with
the equation for an one-dimensional hydrogen atom, whereas 
the quantized energy levels are exactly look like that of the energy 
levels for an
one-dimensional quantum harmonic oscillator. 

Therefore we may conclude by saying that in an uniformly accelerated
frame or in Rindler space, the Schr$\ddot{\rm{o}}$dinger equation for a
particle reduces to the identical form of differential equation satisfied by an one-dimensional hydrogen atom.
Further, the restriction imposed on the solution, to make it physically 
acceptable, gives the quantized energy levels.
The energy levels are found to be exactly identical with that of one-dimensional quantum harmonic oscillator. The
energy levels vary linearly with the quantum number $n$, instead of $1/n^2$, where the last one is the case for an
one-dimensional hydrogen atom. The
energy levels are observed to be independent of particle rest mass. 
It depends only on the acceleration $\alpha$ of the frame.
It is evident from eqn.(22) that $\omega \longrightarrow 0$ as $\alpha \longrightarrow 0$.
Further, unlike the one-dimensional quantum mechanical 
harmonic oscillator, the minimum energy of the particle or the ground state energy for a
given $\alpha$ is
$\hbar \omega$ for $n=1$, i.e., there is no zero point energy. It is
also obvious from this analysis that the energy levels are produced by the
uniform gravitational field or the constant acceleration of the frame. Therefore if any transition takes place from some higher to lower energy
levels, the emitted energy will not be of any kind of conventional or known type quanta.
We call it as the cosmic phonon. In the case of excitation to some
higher energy levels the absorbed energy must also be in the form of cosmic phonon. 

Now writing $ \omega = \frac{\alpha}{c}= 2\pi \nu$, we have
$\nu=\frac{\alpha}{2\pi c}$. 
Hence writing $ \nu \lambda =c$, where $\lambda$ is the cosmic phonon wavelength and assuming that these quanta are
also traveling with the speed of light, we have 
$$\lambda \alpha = 2\pi c^2 = {\rm{constant}}\eqno(23a) $$
This is equivalent to
gravitational Wein's displacement law,
whereas for the usual black body case it is given by
\begin{equation}
\lambda
T={\rm{constant}}
\end{equation}
where $T$ is the temperature of the black body
system. To elaborate this point a little more, we assume that if a large
number of cosmic
phonons are created in the very early universe, 
within a region where the gravitational field was uniform, 
and are created even before the epoch
when the matter and radiation are not decoupled, 
then as the universe expands, since the gravitational field
decreases, the wave length of this non-thermal cosmic phonon 
field will increase.
Therefore, the gravitational field $\alpha$ plays the role of $T$, the equilibrium temperature 
for
the thermal field, e.g., the CMBR. Therefore based on
our model calculation we may assume that the non-thermal cosmic phonon
field, which may also be assumed to be some kind of neutral scalar field
and the thermal field CMBR may exist side by side.
Since the energy levels are created in presence of a
background uniform gravitational field, the concept of spin of the
emitted or absorbed cosmic phonons can not be predicted here. Again defining
some kind of refractive index $\mu \propto \alpha$ \cite{R1}, we have 
$$\mu
\lambda ={\rm{constant}} \eqno(23b) $$
Since the cosmic gravitational field or the inter-galactic gravitational field at the
present epoch is low enough, assuming Newtonian form of gravitational
field, we may write 
\begin{equation}
\alpha \propto \frac{1}{x_l^2}
\end{equation}
where $x_l$ is the position of the local rest frame, in which within a limited region the gravitational field is
assumed to constant. Then
\begin{equation}
\lambda\propto x_l^2
\end{equation}
Therefore the wavelength decreases as the square of some length
parameter.
Further, it is quite possible that an enormous number of such cosmic
phonons might have produce at the proximity of supper massive black
holes present at
the centre of the galaxies. However, in this case since there is no effective 
expansion of the galaxy, these cosmic phonons or the dark quanta remain confined at the vicinity of massive black
holes. Therefore it is quite likely that these quanta could be treated as one of the viable candidates for
non-baryonic dark matter and play active role in galaxy formation.
Of course at present we can not give any experimental technique to
detect these cosmic phonons as dark matter. Further, the emission of
cosmic phonons in Rindler space are because of vacuum excitation by
the accelerted particles. Hence it may be assumed
to be some kind of Unruh process \cite{UN1,UN2}.
\section{Schr\"{o}dinger Equation in Rindler Space- A Linear
Approximation}
In the non-relativistic approximation, i.e. for $m_0 c^2 \gg pc$, the 
Hamiltonian given by eqn.(9) reduces to
\begin{eqnarray}
H&\approx& m_0c^2\left (1+\frac{\alpha x}{c^2}\right ) \left (
1+\frac{p^2}{2m_0^2c^2}\right )\nonumber \\
&=&
\left (1+\frac{\alpha x}{c^2}\right ) \left (
m_0c^2+\frac{p^2}{2m_0}\right )\nonumber \\
\end{eqnarray}
In this approximation, the Schr$\ddot{\rm{o}}$dinger equation for the particle is then given by
\begin{equation}
H\psi =\left ( 1+\frac{\alpha x}{c^2} \right ) \left ( m_0c^2
+\frac{p^2}{2m_0} \right )\psi =E\psi
\end{equation}
Then using the representation
\begin{equation}
\frac{p^2}{2m_0}=-\frac{\hbar^2}{2m_0} \left (
\frac{\partial^2}{\partial x^2} +\frac{\partial^2}{\partial y^2}
+\frac{\partial^2}{\partial z^2} \right )
\end{equation}
we have after a little algebraic manipulation of eqn.(27)
\begin{eqnarray}
-\frac{\hbar^2}{2m_0} \left (
\frac{\partial^2}{\partial x^2} +\frac{\partial^2}{\partial y^2}
+\frac{\partial^2}{\partial z^2} \right ) \psi(x,y,z)&+& \frac{\alpha
Ex}{c^2}\psi \\ \nonumber  &=&E_k\psi
\end{eqnarray}
where the kinetic energy of the particle $E_k=E-m_0c^2$. Then it is quite
obvious that in the separable of variables form (with eqn.(12)), the above equation
may be written as 
\begin{equation}
\frac{d^2 X}{d x^2}-\frac{2m_0 E\alpha}{\hbar^2 c^2}x
X(x) =-\frac{2m_0}{\hbar^2} \left ( E_k- \frac{p_\perp^2}{2m_0}
\right ) X(x)
\end{equation}
where
\begin{equation}
\frac{p_\perp^2}{2m_0}= \frac{p_y^2+p_z^2}{2m_0}
\end{equation}
is the orthogonal part of kinetic energy. Hence the parallel
part of kinetic energy is given by
\begin{equation}
E_{\vert\vert}=E_k-\frac{p_\perp^2}{2m_0}
\end{equation}
Let us put 
\begin{equation}
\zeta=\left ( \frac{2m_0E\alpha}{\hbar^2c^2}\right )^{1/3}x
\end{equation}
a new dimensionless variable and 
\begin{equation}
E^\prime=\frac{2m_0E_{\vert\vert}}{\hbar^2}\left (
\frac{\hbar^2c^2}{2m_0E\alpha}\right )^{2/3}
\end{equation}
as another dimensionless quantity.
Then it can very easily be shown that with $\xi=E^\prime-\zeta$, 
the above differential equation (eqn.(30)) reduces to
\begin{equation}
\frac{d^2 X}{d\xi^2}+\xi X=0
\end{equation}
This equation is of the same form as was obtained by Fowler and
Nordheim in their original work on field emission of electrons \cite{FN} 
(see equation before eqn.(7) in \cite{FN}). The identical mathematical
 structure of the differential equations have come from the same kind of constant driving fields in both the cases,
 i.e. from the identical type of physical reasons. 
In the case of Fowler-Nordheim emission, it is
the constant attractive electrostatic field derived from the triangular type potential 
of the form $C-eEx$,
where $C$ is the surface barrier, which is approximated with the work
function of the metal, $E$ is the uniform electrostatic field
near the metal surface and $e$ is the magnitude of electron charge.
The quantity $C-eEx$ acts as the driving 
potential for cold emission. Whereas in the
case of black hole emission the driving force 
is the uniform gravitational field near the event horizon of
the black hole \cite{HW1,HW2}.

In the {\bf{A2}} part of the Appendix we have given an outline to obtain the solution of the 
differential equation given by eqn.(35). With this solution, we have
\begin{eqnarray}
\psi(x,y,z)&=&N\exp\left (-i\frac{p_yy}{\hbar}\right )\exp\left (-
\frac{ip_zz}{\hbar}\right )\nonumber \\ &&
(E^\prime-\zeta)^{1/2}H_{1/3}^{(2)}\left [
\frac{2}{3}(E^\prime-\zeta)^{3/2}\right ]
\end{eqnarray}
where $N$ is the normalization constant.
Since we expect oscillatory solution along $x$-direction also in the 
asymptotic 
region, where the particles are moving freely, we have 
replaced $J_{1/3}(x)$ by
$H_{1/3}^{(2)}(x)$, the Hankel function of second kind. Now, from the previous 
definitions
\begin{equation}
\xi=E^\prime -\zeta=\frac{2m_0E_{\mid\mid}}{\hbar^2} \left (
\frac{\hbar^2c^2}{2m_0E\alpha}\right )^{2/3}-\left ( \frac{2m_0E\alpha}
{\hbar^2/c^2} \right )^{1/3} x
\end{equation}
if it is assumed that for some local rest frame at a distance $x_l$ from the centre of the black hole, 
in the asymptotic region, where the particles are moving freely,
i.e., $x_l \gg$ the Schwarzschild  radius, the strength of gravitational field
$\alpha=GM/x_l^2$, the  quantity $\xi$ as defined above can be
expressed in terms of $x_l$ in the following manner.
\begin{equation}
\xi \sim ax_l^{4/3}-bx_l^{1/3}
\end{equation}
where $a$ and $b$ are real positive constants.
The argument of the Hankel function, (which in the present
physical scenario is the 
appropriate solution for the differential equation, given by
eqn.(35)) is large enough and positive in the asymptotic region. The
Hankel function can therefore be expressed as an oscillatory function in this uniformly 
accelerated frame in the asymptotic region. This is to be noted that
here we are not talking about the variation of $\alpha$. It is a constant 
for a particular frame of reference, called local frame, having
spatial coordinate $x_l$, or equivalently for a frame at rest in 
presence of an uniform gravitational field $\alpha$, known as local acceleration.
To make this point more transparent, we have considered a large 
number of uniformly accelerated frame of references
in the space out side a black hole, situated near close 
proximity of event horizon to asymptotically far away from
it. Each of these frames are designated by the 
spatial coordinate $x_l$ in one dimension along the positive $x$-direction, measured from
the centre of the black hole. However, the problem we are dealing here has positive-negative symmetry in
$x$-coordinate. Here to keep one to one 
correspondence with Fowler-Nordheim field emission, we have assumed 
one dimensional configuration with motion along positive
$x$=direction.

On the other hand if it is assumed that the uniform acceleration for
a local frame at $x_l$, close to the event horizon, is blue shifted, or in
other words the gravitational field is assumed to be blue shifted for
a local frame at rest at $x_l$ near the event horizon, one can write
\begin{equation}
\alpha=\frac{GM}{x_l^2} \left [ 1-\frac{R_s}{x_l}\right ]^{-1/2}
\end{equation}
which gives the diverging value for $\alpha$ at the Schwarzschild radius, 
i.e., for
$x_l=R_s=2GM/c^2$, or equivalently speaking, if the uniformly accelerated frame is 
considered exactly at the event horizon. 
It should be noted that the value of $\xi$ is negative (from eqn.(38)) near the event
horizon and remains negative up to a certain value of $x$ for the local 
rest frames for which $\alpha$'s are quite
large in magnitude. To accommodate the negative values for $\xi$ for a set of 
local rest frames, we
make the following changes in the wave function in the negative
$\xi$ region.
We replace $\xi$ by $-\xi$, and then the modified form of
Hankel function is given by 
\begin{equation}
H_{1/3}^{(2)}\left(\exp\left (\frac{3}{2}\pi i \right  )Q\right
)
\end{equation}
which may be expressed in terms of the modified Bessel function of first
kind and is given by
\begin{equation}
-\frac{1}{\sin(\pi/3)}\left[ I_{-1/3}(Q)+\exp(i \pi/3)I_{1/3}(Q)
\right ]
\end{equation}
where $Q=2\xi^{3/2}/3$. 

Now  we define the particle density in the following manner in a 
particular local rest frame in
presence of gravitational field $\alpha$.
\begin{equation}
n={\rm{constant}} \mid \psi \mid^2
\end{equation}
The number density
will be large enough for the local rest frames near the event horizon 
where $\xi$'s are
negative and are of extremely large in magnitude. This also follows from the expression for modified Bessel
function of first kind for large $Q$ as given below 
\begin{equation}
I_\nu(Q)\sim \frac{1}{(2\pi Q)^{1/2}} \exp(Q)
\end{equation}
The physical reason for large particle number density near the
event horizon is due to the strong gravitational field, which
produces more particles compared to far regions. This is also true in
the case of Fowler-Nordheim field emission. The more strong the
electrostatic field more will be the electron emission rate. This may sometime causes the accumulation of space
charge near the metal surface.
Now it can very easily be shown that in this region 
the number density is given by
\begin{equation}
n\sim \xi^{1/2} \exp(2Q)
\end{equation}
Of course the model is not valid exactly at the event
horizon. 

When $\xi$ becomes positive, which is true for a frame quite far away 
from the
event horizon, the wave function is given by the Hankel function. 

At $\xi=0$, although the Hankel function diverges, the wave function
vanishes in this particular frame of reference because of 
$\xi^{1/2}$ term. It can very easily be shown that the solution
for $\xi <0$, matches exactly with $\xi>0$ solution at $\xi=0$.
Further the Hankel function asymptotically becomes oscillatory (exponential with imaginary argument) in
nature. The wave function for $\xi\longrightarrow \infty$ is given by
\begin{equation}
\psi(\xi)\sim \xi^{-1/4}\exp\left [ -i\left (\xi -\frac{5\pi}{12}
\right ) \right ]
\end{equation}
Then  the particle density in some local rest frame at $x_l$, which is
far away from the event horizon, in presence of 
an uniform weak gravitational field is given by
\begin{equation}
n(\xi \longrightarrow \infty) \sim (ax_l^{4/3}-bx_l^{1/3})^{-1/2}
\end{equation}

The value of
$\xi=0$ gives $x_l=(E_{\mid\mid}/E)(c^2/\alpha)$, the spatial coordinate 
of a local rest frame where the particle
density is exactly zero. If it is further
assumed that $E_{\mid\mid}=E$, then $x_l=c^2/\alpha$. Therefore the
coordinate point where $\xi$ switches over from negative value to
positive value, depends on the acceleration of the local frame.
Therefore we may divide the whole
space out side the black hole into effectively six regions: 
for the set of local rest frames in presence of 
uniform gravitational field, 
but far from the event horizon, 
the wave functions are oscillatory; for $\xi>0$
but not large enough, the wave functions can be expressed in those
frames in terms of
Hankel function of second kind; at $\xi=0$, the nature of the wave
functions from both $\xi \longrightarrow 0_+$ and $\xi \longrightarrow
0_-$ show that it should vanish; for $\xi <0$, but the magnitude is
not large enough, the wave functions can be expressed in terms of
modified Bessel function of first kind; very close to the event
horizon, where $\xi$ is also less than zero but have very large 
magnitude, the number density shows exponential growth and asymptotically diverges. Finally nothing
can be said at and inside the event horizon. We further conclude that there are particle creation up to $\xi=0$,
beyond which the created particles are moving along $x$-direction and finally far from the event horizon they become
asymptotically free.
\section{Schr\"{o}dinger Equation in Rindler Space- A Quadratic
Approximation}
\noindent Keeping only three terms of the binomial expansion of the factor
$\left( 1 + \frac{\alpha x}{c^2}\right)^{-1}$, the Schr\"{o}dinger equation reduces to 
\begin{eqnarray}
\left(m_0 ^2 c^2 + \frac{p^2}{2m_0}\right)\Psi(x,y,z) &=& \left(1
+\frac{\alpha x}{c^2}\right)^{-1} E \Psi(x,y,z)\nonumber \\
&\approx& \left( 1 -
\frac{\alpha x}{c^2}+\frac{\alpha^2 x^2}{c^4}\right) E \Psi(x,y,z)
\end{eqnarray}
After rearranging this equation we may be written as
\begin{equation}
\frac{p^2}{2m}\Psi(x,y,z) + \frac{\alpha x }{c^2} E\Psi(x,y,z)
-\frac{\alpha ^2 x^2}{c^4}E \Psi(x,y,z) = E_k \Psi(x,y,z)
\end{equation}
 where $ E_k = E- m_0 c^2 $, the kinetic energy of the particle. Now
 substituting the separable form of the wave function $\Psi(x,y,z)$ (eqn.(12)), we have
\begin{equation}
-\frac{\hbar^2}{2m_0}\frac{d^2 X}{dx^2} + \frac{\alpha x }{c^2} E X(x)
-\frac{\alpha ^2 x^2}{c^4}E X(x) = \left(E_k - \frac{p_\perp ^2}{2m_0} \right)X(x)
\end{equation}
where $ p_\perp ^2 = p_y ^2 + p_z ^2 $ . Rearranging and using $
E_{k\parallel} = E_k - \frac{p_\perp ^2}{2m_0}$ the parallel part of the particle kinetic energy, the above differential
equation may be written in the form
\begin{equation}
\frac{d^2 X}{dx^2} + \frac{2m_0 E}{\hbar^2} \left[\frac{\alpha ^2 x^2
}{c^4} -\frac{\alpha x}{c^2}\right] X(x) = - \frac{2m_0 }{\hbar
^2}E_{k\parallel} X(x)
\end{equation}
which may further be expressed in the form
\begin{equation}
\frac{d^2 X}{dx^2} + \frac{2m_0 E}{\hbar^2} \left[\left(\frac{\alpha
x}{c^2}-\frac{1}{2}\right)^2 -\frac{1}{4}\right] X(x) = - \frac{2m_0 }{\hbar
^2}E_{k\parallel} X(x)
\end{equation}
Now changing the variable from $x ~~{\rm{to}} ~~p $, where
 $ p = \frac{\alpha x }{c^2} - \frac{1}{2}$, the above differential equation can be written as
\begin{eqnarray}
\frac{d^2 X}{dp^2} &+&\frac{q^2}{4}\left( p^2 - \frac{1}{4}\right) X +
\frac{q^2}{4}\gamma X = 0\\
&~~{\rm{where}} ~~& q^2 = \frac{8m_0 E c^4}{\alpha ^2 \hbar ^2} 
\end{eqnarray}
and $\gamma = \frac{E_{k\parallel}}{E}$ which is $\leq 1$.
 It should be noted that the variable $ p$ is $ PT$ invariant.
Using new variable $\rho = pq^{1/2}$, we have from the above differential equation 
\begin{equation}
\frac{d^2 X}{d\rho^2} +\left(\frac{\rho^2}{4}-\lambda \right) X(\rho) = 0
\end{equation}
where  $\lambda= \frac{q}{4}\left(\frac{1}{4}-\gamma \right)$.
Obviously $\lambda =0$, $>0$ or $<0$ for $\gamma=1/4$, $\gamma<1/4$ or $\gamma>1/4$ respectively.
Now for $\gamma \longrightarrow1$, that is for the extreme case $E_{k_\parallel}=E$ and $\lambda=-0.75$, the minimum value of
$\lambda$. Whereas for $\gamma \ll 1/4$, i.e., $ E_{k_\parallel} \ll E$, $\lambda=0.25$, the maximum value of $\lambda$. Both are in units of $q/4$.
It can very easily be shown that for purely one dimensional case the energy eigen value $E= \frac{m_0 c^2}{1 - \gamma}$. For $ \gamma =\frac{1}{4}$, the energy
eigenvalue $ E =\frac{4}{3} m_0 c^2$, whereas for $\gamma <\frac{1}{4}$, $E <\frac{4}{3}m_0 c^2$ and for $\gamma >
\frac{1}{4}$, $ E > \frac{4}{3}m_0 c^2$. Since the energy is always
finite, we should have $ \gamma <1$, which is also obvious from the
definition of $\gamma$. Hence one can very easily
show that 
\begin{equation}
\frac{q}{4} \approx \frac{m_0 c^2}{(1-\gamma)\hbar \omega} 
\end{equation}
where we have put $2^{1/2} \approx 1$ and $\omega = \alpha /c$, some kind of frequency. 
The solution of the above
differential equation for $\lambda\neq 0$ is the parabolic cylindrical function $W(\lambda,\pm \rho)$, given by
\cite{R11}
\begin{equation}
W(\lambda,\rho)=W(\lambda,0)\omega_1(\lambda,\rho)+W^\prime(\lambda,0)\omega_2(\lambda,\rho)
\end{equation}
where 
\begin{eqnarray}
W(\lambda,0)&=&2^{-3/4}\left | \frac{\Gamma\left (\frac{1}{4}+\frac{1}{2}i\lambda \right )} 
{\Gamma\left (\frac{3}{4}+\frac{1}{4}i\lambda \right )} \right |^{1/2}~~{\rm{and}} \nonumber \\
W^\prime(\lambda,0)&=&-2^{-1/4}\left | \frac{\Gamma\left (\frac{3}{4}+\frac{1}{2}i\lambda \right )} 
{\Gamma\left (\frac{1}{4}+\frac{1}{4}i\lambda \right )} \right
|^{1/2}
\end{eqnarray}
\begin{eqnarray}
\omega_1(\lambda,\rho)&=&\sum_{n=0}^\infty\alpha_n(\lambda)
\frac{\rho^{2n}}{2n!} ~~{\rm{and}} \nonumber \\
\omega_2(\lambda,\rho)&=&\sum_{n=0}^\infty\beta_n(\lambda)
\frac{\rho^{2n+1}}{(2n+1)!}
\end{eqnarray}
where $\alpha_n(\lambda)$ and $\beta_n(\lambda)$ satisfy the recursion
relations
\begin{eqnarray}
\alpha_{n+2}&=&\lambda \alpha_{n+1}-\frac{1}{2} (n+1)(2n+1)\alpha_n,
\nonumber \\
\beta_{n+2}&=&\lambda \beta_{n+1}-\frac{1}{2}
(n+1)(2n+3)\beta_n ~~{\rm{and}}~~ \nonumber \\ 
\alpha_0(\lambda)&=&\beta_0(\lambda)=1, ~~ \alpha_1(\lambda) =
\beta_1(\lambda)=\lambda
\end{eqnarray}
To show the variation of the wave function with $\rho$, in fig.(1) we have plotted $\mid W(\rho,\lambda)\mid^2$, 
the probability density, against $\rho$, for $\lambda=0.5$, curve (a),
$0.25$, curve (b), $-0.05$, curve (c), $-0.5$, curve (d) and $-0.75$,
curve (e) in
units of $q/4$. In the numerical evaluation of $W(\rho,\lambda)$, we have obtained the absolute values for the
ratio of $\Gamma$-functions using the formula \cite{R11}
\begin{equation}
\left |\frac{\Gamma(a+ib)}{\Gamma(a)}\right |^2=\prod^{\infty}_{n=0}\left[1+\frac{b^2}{(a+n)^2}\right]^{-1}
\end{equation}
where both $a$ and $b$ are real constants. One can conclude from the nature of the curves that the probability
densities are of damped oscillatory in nature.

Now for the special case with $\lambda=0$, the differential equation given by eqn.(54) reduces to
\begin{equation}
\frac{d^2X}{d\rho^2}+\frac{\rho^2}{4}X(\rho)=0
\end{equation}
With a new variable $u=\rho/\sqrt2$, this equation becomes
\begin{equation}
\frac{d^2X}{du^2}+u^2X=0
\end{equation}
In {\bf{A3}} part of the Appendix, we have obtained the solution of this differential equation and is given by
\begin{equation}
X(\rho)=\frac{\rho^{1/2}}{2^{1/4}}J_{1/4}\left(\frac{\rho^2}{4}\right)
\end{equation}
for $\rho \geq0$. Here $J_n(x)$ is the Bessel function of order n. In fig.(2) we have shown the variation of
$\mid X(\rho)\mid^2$ with $\rho$. The nature of the probability density $\mid X(\rho)\mid ^2$ is also damped
oscillatory type. In this case $\gamma = \frac{1}{4}$, $E=\frac{4}{3}m_0 c^2$ and $\frac{q}{4}$ is exactly equal to
$\frac{m_0 c^2}{\hbar \omega}$.
\section{Exact Solution}
In the present formalism we use natural units, $\hbar = c=1$. The Hamiltonian in the Rindler space is then given by
\begin{equation}
H = (1+\alpha x)(p^2 +m_0^2)^{1/2}
\end{equation}
Then the Schr\"{o}dinger equation $H\Psi=E\Psi$ may be written as 
\begin{equation}
(1 +\alpha x)(-d_x^2 +m_0^2)^{1/2}\Psi = E \psi
\end{equation}
where $ d_x = \frac{d}{dx}$ and we assume that the motion is one dimensional and along
positive $ x- $direction.
Changing the variable from $ x ~~{\rm{to}}~~ X $, given by
$X = 1 +\alpha x $, 
the above equation reduces to 
\begin{equation}
X(-d_X^2 +m^{*2})^{1/2} \Psi = E^*\psi
\end{equation}
where $m^*= m_0/\alpha ~~{\rm{and}}~~ E^*= E/\alpha.$ In the text
below we shall reset $ m^*= m_0
~~{\rm{and}}~~ E^* = E.$ Then we can rewrite the above differential equation as
\begin{equation}
X(-d_X^2 +m_0 ^2)^{1/2} \Psi = E\psi
\end{equation}
To get an analytical solution, we follow the technique presented in \cite{R10a,R10b,R10c}.
Now using the properties of Dirac delta function, we can write \cite{R10a,R10b,R10c} the left hand side of the
above equation in the form
\begin{equation}
X (-d_X^2 +m_0^2)^{1/2} \Psi(X) = \int_{-\infty}^{+\infty}
X(-d_X^2 +m_0^2)^{1/2}\delta(q-X) \Psi(X) dq
\end{equation}
 Since $\delta(x-a)f(x) = \delta(x-a)f(a)$, we have
\begin{equation}
X (-d_X^2 +m_0^2)^{1/2} \Psi(X)  = \int_{-\infty}^{+\infty}
q(-d_X^2 +m_0^2)^{\frac{1}{2}}\delta(q-X) \Psi(q) dq
\end{equation}
Using the integral representation of $ \delta -$function, given by
\begin{equation}
\delta(q-X) = \frac{1}{2\pi}\int_{-\infty}^{+\infty} dp \exp[-i(q-X)p]
\end{equation}
we have
\begin{eqnarray}
X (-d_X^2 +m_0^2)^{1/2} \Psi(X)  &=& \frac{1}{2\pi}\int_{-\infty}^{+\infty}
\int_{-\infty}^{+\infty} q(p^2 +m_0^2)^{1/2}\nonumber \\
&&\exp[-i(q-X)p] \Psi(q)
dp dq\nonumber \\
\end{eqnarray}
Hence we can re-write the left hand side in the the following form
\begin{eqnarray}
X (-d_X^2 +m_0^2)^{1/2} \Psi(X) = 
\frac{1}{2\pi}(-d_X^2+m_0^2)&&\int_{-\infty}^{+\infty} q\Psi(q)dq
\nonumber \\
\int_{-\infty}^{+\infty}dp\frac{\exp[-i(q-X)p]}{(p^2 +m_0^2)^{1/2}} 
\end{eqnarray}
with some simple algebraic manipulation, the above expression is given by \cite{R11}
\begin{eqnarray}
X (-d_X^2 +m_0^2)^{\frac{1}{2}} \Psi(X) & = &
\frac{1}{\pi}(-d_X^2+m_0^2)\int_{-\infty}^{+\infty} q\Psi(q)dq
\int_{0}^{\infty}dp\frac{\cos[(q-X)p]}{(p^2 +m_0^2)^{\frac{1}{2}}}
\nonumber \\ & = & 
\frac{1}{\pi}(-d_X^2 +m_0^2)\int_{-\infty}^{+\infty} dq q \Psi(q)
K_0(m_0|q-X|)
\end{eqnarray}
Dividing the $q$ integral into two parts, we have
\begin{eqnarray}
&&X (-d_X^2 +m_0^2)^{\frac{1}{2}} \Psi(X)  =   
 \frac{1}{\pi}(-d_X^2 +m_0^2)\times \nonumber \\ &&
 \left[\int_{-\infty}^{X} dq q \Psi(q)
K_0[m_0(X-q)] +\int_{X}^{+\infty} dq q \Psi(q) K_0[m_0 (q-X)]\right]
\end{eqnarray}
Then substituting $ X-q = q_1 $ in the first integral and $ q-X = q_2 $ in
the second integral and redefining $ q_1 = q$ in the first integral, 
and $q_2 =q$ in the second integral,
we have 
\begin{eqnarray}
&&X (-d_X^2 +m_0^2)^{\frac{1}{2}} \Psi(X)  =
\frac{1}{\pi}(-d_X^2 +m_0^2)\times \nonumber \\
&&\int_{0}^{\infty} dq
K_0(m_0 q) \left[ (X+q) \Psi(X+q) +(X-q)\Psi(X-q)\right]
\end{eqnarray}
We seek the series solution
\begin{equation}
\Psi(X) = \sum_{k=1}^{n+1} \gamma_{k,n+1} X^k \exp(-\beta X)
\end{equation}
 where $\gamma_{k,n+1} ~~{\rm{and}}~~ \beta $ are 
 unknown constants, to be obtained
 from the recursion relations.  
Then the $n_{th}$ term is given by
\begin{eqnarray}
&&X (-d_X^2 +m_0^2)^{1/2} X^n \exp(-\beta X) =
\frac{1}{\pi}(-d_X^2 +m_0^2)\exp(-\beta X)\times \nonumber \\
&&\int_{0}^{\infty} dq
K_0(m_0 q) \left[ (X+q)^{n+1} \exp(-\beta q) +\exp(\beta q)
(X-q)^{n+1}\right] 
\end{eqnarray}
Now expanding $(X+q)^{n+1}$ and $(X-q)^{n+1}$ in Binomial series and then using the standard relation we have
\cite{R11}
\begin{eqnarray}
&&\int_{0}^{\infty} x^{\mu -1} \exp(-\alpha x) K_{\nu}(\beta_1 x
) dx = \nonumber \\
&& \frac{(\pi)^{\frac{1}{2}} (2\beta_1) ^\nu}{(\alpha +\beta_1) ^{\mu
+\nu}}\frac{\Gamma(\mu +\nu)\Gamma(\mu - \nu)}{\Gamma(\mu
+\frac{1}{2})}
F\left(\mu +\nu , \nu +\frac{1}{2}; \mu +\frac{1}{2}; \frac{\alpha -
\beta_1}{\alpha + \beta1}\right) 
\end{eqnarray}
where $F(a,b;c;d) $ is the Hypergeometric function and in our case with $ \nu =0
,~ x=q ,~ \mu -1 = k+1 ,~ \alpha = \beta ~~{\rm{and}}~~ 
\beta_1 =m_0 $, the
integral reduces  to
\begin{equation}
I = \frac{\pi ^{1/2} [\Gamma(k+2)]^2}{\Gamma(k+\frac{5}{2})
(m_0+\beta)^{k+2}}F\left(k+2, \frac{1}{2} ; k+\frac{5}{2} ; -\frac{m_0-
\beta}{m_0 +\beta}\right)
\end{equation}
Then after a little algebra, we have
\begin{eqnarray}
&&{X (-d_X^2 +m_0^2)^{1/2} X^n \exp(-\beta X) = } \nonumber \\
&&\frac{1}{\pi^{1/2}}(-d_X^2 +m_0^2)\exp(-\beta X)\sum_{k=0}^{n+1}
\left(\begin{array} {c} n+1 \\ k \end{array} \right) G_k(m_0,
\beta) X^{n+1-k} 
\end{eqnarray}
where
\begin{eqnarray}
&&G_k(m_0 , \beta) =
\frac{[\Gamma(k+2)]^2}{\Gamma(k+\frac{5}{2})} 
\left[\frac{1}{(m_0+\beta)^{k+2}}F\left(k+2, \frac{1}{2} ; k+\frac{5}{2}
; -\frac{m_0-\beta}{m_0 +\beta}\right)\right ] +\nonumber \\ 
&&\frac{[\Gamma(k+2)]^2}{\Gamma(k+\frac{5}{2})} 
 (-1)^k \left [\frac{1}{(m_0-\beta)^{k+2}}F\left(k+2,\frac{1}{2} ;
k+\frac{5}{2} ; -\frac{m_0+\beta}{m_0 -\beta}\right) \right ]
\end{eqnarray}
Now it is a matter of simple algebra to show that
\begin{eqnarray}
&&X (-d_X^2 +m_0^2) X^{n+1-k} \exp(-\beta X) = \nonumber \\
&&(m_0^2 - \beta ^2)
\exp(-\beta X) X^{n+1-k} + 2(n+1-k) \beta \exp(-\beta X) X^{n-k}
\nonumber \\
&-& (n+1-k)(n-k) \exp(-\beta X) X^{n-k-1}
\end{eqnarray}
Then
\begin{eqnarray}
&&X (-d_X^2 +m_0^2)  X^n \exp(-\beta X) = \nonumber \\
& &\frac{1}{\pi^{1/2}}(-d_X^2 +m_0^2)\exp(-\beta X)\sum_{k=0}^{n+1}
\left(\begin{array} {c} n+1 \\ k \end{array} \right) \nonumber \\
& &\left[(m_0^2 - \beta ^2)G_k(m_0 , \beta) + 2\beta k G_{k-1}(m_0 , \beta)
-k(k-1)G_{k-2}(m_0 , \beta)\right] X^{n+1-k}  \nonumber \\
\end{eqnarray}
Hence from the quantum mechanical equation
\begin{equation}
X (-d_X^2 +m_0^2)^{1/2} \Psi (X) = E\Psi(X) ,
\end{equation}
we have from the series solution (polynomial form) of $ \Psi(X) $,
\begin{eqnarray}
\lefteqn{X(-d_X^2 +m_0^2)^{1/2} \sum_{k=1}^{n+1} \gamma_{k,n+1} X^k \exp(-\beta
X) }\nonumber \\
& = &\sum_{k=1}^{n+1} \gamma_{k,n+1} \sum_{p=0}^{k} F_{p,k}(m_0,\beta)
X^{k+1-p}  \nonumber \\
& = & E \sum_{k=1}^{n+1} \gamma_{k,n+1}\exp(-\beta X) X^k
\end{eqnarray}
The expressions for $F_{p,k} (m_0,\beta)$ 
and $\gamma_{k,n+1}$ have been derived in the {\bf{A4}} part of the Appendix
in terms the parameters $\beta$ and the rest mass $m_0$.
From the above equation equating the coefficient of $x^2$, we have
\begin{equation}
E_n = \frac{\gamma_{1,n+1}}{\gamma_{2,n+1}} F_{0,n+1}(m_0 ,\beta)
\end{equation}
the energy corresponding to the $n_{th}$ level of the spectrum. 
Further equating the coefficients of $X^l$ from both the sides and
putting $p=1$, we have
\begin{eqnarray}
&&\gamma_{l,n+1}(\beta,m_0) F_{1,l}(m_0 , \beta) = E_l(\beta,m_0) \gamma_{l,n+1}(m_0,\beta) ~~{\rm{or}} \\
&& E_l(m_0,\beta) = F_{1,l} (m_0 ,\beta)
\end{eqnarray}
which gives the energy spectrum, provided the parameter $\beta$ is known.
Where
\begin{equation}
F_{1,l}(m_0,\beta)=\frac{-l\beta}{(2\pi m_0)^{1/2}(m_0 ^2 - \beta
^2)^{3/4}} P_{-\frac{1}{2}}^{-\frac{3}{2}}\left
(\frac{\beta}{m_0}\right ) \\
\end{equation}
where $l=1,2,3,......$, positive integers.
Therefore to obtain the energy spectrum, we have to evaluate the
associated Legendre function. It is to be noted further that the energy
spectrum is real and linearly quantized. The wave functions are
bounded (becaus of the factor $\exp(-\beta X)$). From the above
expression it is quite obvious that $\beta < m_0$. This is also a
necessary condition for the argument $z$ of $P_\nu^\mu(z)$.
In fig.(3) we have plotted the variation of $E_l$ with
$\beta$ (the parameter has been re-defined as $\beta/m_0$) for $l=1$. 
Since the energy levels are proportional to $l$ we
have not considered other $l$-values. We have noticed that the magnitude of the energy eigen 
value $\mid E_l(m_0,\beta)\mid $ increases with $\beta$ and the rise is
very sharp as $\beta \longrightarrow 1$. However it is negative and the negativity increases with $\beta$. Further,
the eigen functions are $\propto \exp(-\beta X)$, therefore with the increase of $\beta$ the wave function converges
to zero very quickly, whereas the eigen states become more bound
because of high negative value of energy, or in other words, because
of very high binding energy.
Therefore with the increase of negative value of the energy makes the state more bound and simultaneously the
spread of wave function in space decreases. The later is also in agreement with more stronger binding.
\section{Klein -Gordon Equation }
To obtain the modified form of Klein-Gordon Equation in a uniformly
accelerated frame we consider the classical Hamiltonian
\begin{equation}
H = \left( 1+\frac{\alpha x}{c^2}\right)(m_0^2 c^4 + p^2c^2)^{\frac{1}{2}}
\end{equation}
Squaring both the sides, we have
\begin{equation}
H^2 = \left( 1+\frac{\alpha x}{c^2}\right)^2(m_0^2 c^4 + p^2c^2)
\end{equation}
which is the classical form of square of the Hamiltonian in Rindler space. 
Then the modified form of the Klein-Gordon equation is given by
\begin{equation}
H^2 \Psi(x,y,z) = E^2 \Psi(x,y,z)
\end{equation}
where $E$ is the energy eigen value.
Writing explicitly the Hamiltonian part and using the separable form of $\Psi(x,y,z) ,$
(eqn.(12)), we have
\begin{equation}
\left( 1 +\frac{\alpha x}{c^2}\right)^2 \left( -\hbar ^2 c^2 \frac{d^2 X}{dx^2} + 
E_\perp ^2 X(x)\right) = E^2 X(x)
\end{equation}
where $E_\perp ^2= (p_y^2+ p_z^2)c^2 + m_0^2 c^4 ,$ the square of the
transverse part of particle energy in the relativistic form. 
Substituting $1+\frac{\alpha x}{c^2} = u , $ which is $PT$ symmetric,
the above equation may be written in the form
\begin{equation}
\frac{ \hbar ^2 \alpha ^2} {c^2} \frac{d^2 X}{du^2} + \frac{E^2}{u^2} X(u)
=E_\perp ^2 X(u)
\end{equation}
which may further be written as 
\begin{equation}
\frac{d^2 X}{du^2} +\frac{a}{u^2} X(u) = bX(u)
\end{equation}
with $a = \frac{c^2 E^2}{\hbar ^2 \alpha ^2}$ and $b = \frac{c^2
E_\perp ^2}{\hbar ^2 \alpha ^2}$. 
Let us put $w = u b^{1/2}$, 
then we have
\begin{equation}
\frac{d^2 X}{dw^2} +\frac{a}{w^2} X(w) - X(w) = 0
\end{equation}
Expressing $X = w^n Y , $ we have 
\begin{equation}
w^n \frac{d^2 Y}{dw^2} + 2n w^{n-1} \frac{dY}{dw} +\left[ {n(n-1) +a}
w^{n-2} - w^n \right] Y(w) = 0
\end{equation}
Putting $ n= \frac{1}{2} $, i.e., $X=w^{1/2}Y$, the above equation reduces to
\begin{equation}
w^2 \frac{d^2 Y}{dw^2} +  w \frac{dY}{dw} +\left[ -w^2 -(\frac{1}{4} -
a)\right] Y(w) = 0
\end{equation}
Further on substituting $w\longrightarrow iw$, we have after rearranging the above equation
\begin{equation}
w^2 \frac{d^2 Y}{dw^2} +  w \frac{dY}{dw} +\left[ w^2 -(\frac{1}{4} -
a)\right] Y(w) = 0
\end{equation}
Which may be written as 
\begin{equation}
w^2 \frac{d^2 Y}{dw^2} +  w \frac{dY}{dw} +\left[w^2 -\nu ^2 \right]
Y(w) = 0 
\end{equation}
where $\nu^2 = \frac{1}{4} -a = \frac{1}{4} - \frac{c^2 E^2}{\hbar ^2
\alpha ^2}$. 
For $\nu = 0 $, 
\begin{equation}
E = \frac{1}{2} \frac{\hbar \alpha}{c}
\end{equation}
Defining $\frac{\alpha}{c} = \omega_0 $, some frequency, we have
\begin{equation}
E = \frac{1}{2} \hbar \omega_0 ,
\end{equation}
the zero point energy in some sense.
The solution is given by
\begin{equation}
Y(w) = J_0(iw)
\end{equation}
The zeroth order Bessel function of purely imaginary argument. Then
\begin{equation}
X(w) = w^{\frac{1}{2}} J_0(iw)
\end{equation}
For $ \nu^2 > 0, ~~{\rm{i.e.,}}~~ ~~ \frac{1}{4} - \frac{c^2 E^2}{\hbar ^2 \alpha ^2} >0$, 
or,
\begin{equation}
E < \frac{1}{2} \hbar \omega_0 ,
\end{equation}
which is less than the zero point energy.Since in quantum mechanics the energy eigen value is always greater than or
equal to zero point energy, therefore we have $\nu ^2 <0$. Hence the differential equation reduces to
\begin{equation}
w^2 \frac{d^2 Y}{dw^2} +  w \frac{dY}{dw} +\left[w^2 +\nu ^2 \right]
Y(w) = 0 
\end{equation}
Putting $\nu\longrightarrow i\nu$, we have again
\begin{equation}
w^2 \frac{d^2 Y}{dw^2} +  w \frac{dY}{dw} +\left( w^2 -\nu^2 \right) Y(w) = 0
\end{equation}
This is the differential equation for Bessel function of order $i\nu$ (imaginary order) and $iw$ (imaginary
argument).
The solution for real order but imaginary argument is  well known and is given by
\begin{equation}
Y(w) = J_\nu(iw) = -\left(\frac{iw}{2}\right)^\nu \sum_{n=0}^{\infty}
(-1)^n \frac{1}{(n!) \Gamma(\nu+\mu +1)}
\left(\frac{w}{2}\right)^{2n} 
\end{equation}
Then
\begin{equation}
X(w) = - \left(\frac{i}{2}\right)^\nu w^{\nu +\frac{1}{2}}
\sum_{n=0}^{\infty} (-1)^n \frac{1}{(n!) \Gamma(\nu+\mu+1)}
\left(\frac{w}{2}\right)^{2n}
\end{equation}
Next we consider $\nu^2 <0, ~~{\rm{i.e.,}}~~ ~~\nu \longrightarrow i\nu ~~{\rm{and}}~~ 
E> \frac{1}{2} \hbar \omega$, then 
\begin{equation}
Y(w) = J_{i\nu}(iw) 
~~{\rm{and}}~~ X(w) = w^{\frac{1}{2}} J_{i\nu}(iw) 
\end{equation}
Since the most acceptable value of lower limit / ground state of energy
spectrum is $\geq \frac{1}{2}\hbar \omega$, therefore, the eigen states
are represented by the Bessel function of both imaginary orders and
imaginary arguments. for the sake of some more physical insight, 
let us make a detail analysis of this solution \cite{MF,GB}.
To get an analytical solution, we start with a series solution of
the form 
\begin{equation}
Y(w) = A(w) \cos(\gamma \ln w) + B(w) \sin (\gamma \ln w)
\end{equation}
where $\gamma$ is some parameter and
\begin{eqnarray}
A(w)&=& a_0 + a_1 w + a_2 w^2 + .....+ a_n w^n +.... \\
B(w) &=& b_0 + b_1 w + b_2 w^2 + ....+ b_n w^n +....
\end{eqnarray}
with $a_0 , a_1 , .... ~~ {\rm{and}} ~~ b_0 , b_1 ,....$ are unknown
parameters and are independent of each other.
We substitute $Y(w)$  given  above (eqn.(117)) in the differential
equation given by eqn.(108). Now equating the coefficient of $w^n$ to zero, we get
\begin{eqnarray}
&&\big[ a_n n(n-1)\cos(\gamma\ln w) + b_n n(n-1)\sin(\gamma\ln w) - 2 a_n n \sin
(\gamma \ln v )\gamma \nonumber \\ &+& 
2 b_n n \cos (\gamma \ln v) \gamma + a_n \sin
(\gamma \ln v) \gamma - b_n \cos (\gamma \ln w)\gamma \nonumber \\
&+& a_n n w \cos
(\gamma \ln w) +b_n n \sin (\gamma \ln w ) - a_n \gamma \sin (\gamma \ln
w) \nonumber \\ &+& b_n \gamma \cos (\gamma \ln w) \big ] w^n  
 -  w^{2+n} \left[ a_n
\cos (\gamma \ln w) + b_n \sin (\gamma \ln w) \right] = 0 
\end{eqnarray}
For $n = 1$
\begin{eqnarray}
&&\sin (\gamma \ln w) \left[-2 a_1 \gamma w + b_1 w - b_1 w^3 \right]
\nonumber \\
&+& \cos (\gamma \ln w) \left[2b_1 w \gamma + a_1 w - a_1 w^3 \right]
= 0
\end{eqnarray}
Hence, since $\sin (\gamma \ln w)$ and $\cos(\gamma \ln w)$ are non-zero, we have 
\begin{eqnarray}
&&-2 a_1 \gamma  + b_1  - b_1 w^2 =0 \\ &&~~{\rm{and}}~~
2b_1  \gamma + a_1  - a_1 w^2 = 0
\end{eqnarray}
These are the simultaneous linear homogeneous equations for $a_1
~~{\rm{and}}~~
a_2$. For the non-trivial solutions of $a_1 ~~{\rm{and}}~~ a_2$, we must have
\begin{equation}
\left| \begin{array} {lr} -2\gamma & 1-\omega^2 \\ 1-\omega^2 & 2\gamma
\end{array} \right| =0
\end{equation}
But $\gamma$ is a non-zero real constant parameter, therefore the above
determinant can not be zero. Hence $ a_1 = b_1 = 0 $, which are the trivial
solutions. 
Now it is a matter of simple algebra to show by equating the coefficient of
$w^n $ to zero,
\begin{eqnarray}
&&a_n = \frac{n a_{n-2} -2 \gamma b_{n-2}}{n(n^2 + 4\gamma ^2)} \\
&&~~{\rm{and}}~~ 
b_n = \frac{n b_{n-2} +2 \gamma a_{n-2}}{n(n^2 + 4\gamma ^2)} 
\end{eqnarray}
Hence it is obvious that the coefficients of all the odd power terms of $w $ in eqns.(112) and (113) are zero, 
i.e., $a_j ~~{\rm{and}}~~ b_j$, where $j$ is odd integer are zero. 
Therefore,
the odd terms will not contribute in the solution.
To get the coefficients of the even power of $w$, we seek the solutions
in the form
\begin{eqnarray}
Y(w) &=& Y_1 (w) + i Y_2 (w)\\ &&~~{\rm{where}}~~ 
Y_1 (w) = C(w) \cos (\gamma \ln iw) + D(w)\sin(\gamma \ln iw) \\
&& ~~{\rm{and}}~~ Y_2 (w) = D(w) \cos (\gamma \ln iw) - 
C(w)\sin(\gamma \ln iw) \\
C(w) &=& \sum_{n=0}^{\infty} C_{2n} \left(\frac{w}{2}\right)^{2n} \\
D(w) &=& \sum_{n=0}^{\infty} D_{2n} \left(\frac{w}{2}\right)^{2n} \\
C_{2n} &=& \frac{n C_{2n-2} - \gamma D_{2n-2}}{n(n^2 + \gamma ^2)} \\ 
&&~~{\rm{and}}~~ 
D_{2n} = \frac{n D_{2n-2} + \gamma C_{2n-2}}{n(n^2 + \gamma ^2)} 
\end{eqnarray}
We use  two sets of $(C_0 , D_0)$. Let us first take $(C_0 , D_0)=
(0,1)$
Then 
\begin{eqnarray}
&&C_2 = - \frac{\gamma}{1 + \gamma ^2}   \\ 
&&D_2 =  \frac{1}{1 + \gamma ^2} \\
&&C_4 = - \frac{3\gamma}{2(1^2 + \gamma ^2)(2^2 + \gamma ^2)} \\
&&D_4 =  \frac{1}{(1^2 + \gamma ^2)(2^2 +\gamma ^2)} - \frac{\gamma ^2}{2(1^2 + \gamma ^2)(2^2 + \gamma ^2)} \\
&&C_6 = - \frac{11 \gamma}{6(1^2 + \gamma ^2)(2^2 + \gamma ^2)(3^2 + \gamma
^2)} + \frac{\gamma ^3}{6(1^2 + \gamma ^2)(2^2 + \gamma ^2)(3^2 + \gamma
^2)} \nonumber \\&& \\
&&D_6 = \frac{1}{(1^2 + \gamma ^2)(2^2 +\gamma ^2)(3^2 + \gamma ^2)} -
\frac{\gamma ^2}{(1^2 + \gamma ^2)(2^2 + \gamma ^2)(3^2 + \gamma ^2)} 
\nonumber \\
\end{eqnarray}
etc., then we can write
\begin{equation}
Y_2(w) =
A1 \cos (\gamma \ln iw)  - B1
\sin (\gamma \ln iw) 
\end{equation}
where 
\begin{eqnarray}
A1&=& 1 + \frac{1}{1+\gamma ^2}\left(\frac{w}{2}\right)^2
+ \left\{ \frac{1}{(1^2 +\gamma ^2)(2^2 + \gamma ^2)} - \frac{\gamma
^2}{2(1^2 + \gamma ^2)(2^2 + \gamma ^2)}\right\} 
\left(\frac{w}{2}\right)^4 \nonumber \\ &+&\left\{ \frac{1}{(1^2 + \gamma ^2)(2^2 +
\gamma ^2)(3^2 + \gamma ^2)} -\frac{\gamma ^2}{(1^2 +\gamma ^2)(2^2
+\gamma ^2)(3^2 +\gamma ^2)} \right\}\left(\frac{w}{2}\right)^6 +
.... \nonumber \\
\end{eqnarray}
and
\begin{eqnarray}
B1&=&-\frac{\gamma}{1^2 +\gamma
^2}\left(\frac{w}{2}\right)^2  - \frac{3\gamma}{2(1^2 +\gamma ^2)(2^2
+\gamma ^2)}\left(\frac{w}{2}\right)^4 \nonumber \\ &+&
\left \{\frac{\gamma ^3}{6(1^2
+\gamma^2)(2^2 +\gamma ^2)(3^2 +\gamma ^2)} -
\frac{\gamma}{6(1^2
+\gamma ^2)(2^2 +\gamma ^2)(3^2 +\gamma ^2)}
\right \}\left(\frac{w}{2}\right)^6 + .... \nonumber \\
\end{eqnarray} 
For the other set $(C_0 , D_0) = (1,0)$, then we have
\begin{eqnarray}
C_2 &=& \frac{1}{1 + \gamma ^2} \\
D_2 &=&  \frac{\gamma}{1 + \gamma ^2} \\
C_4 &=&  \frac{1}{(1^2 + \gamma ^2)(2^2 +\gamma ^2)} - \frac{\gamma ^2}{2(1^2 + \gamma ^2)(2^2 + \gamma ^2)} \\
D_4 &=&  \frac{3\gamma}{2(1^2 + \gamma ^2)(2^2 + \gamma ^2)} \\
C_6 &=&  \frac{1}{(1^2 + \gamma ^2)(2^2 +\gamma ^2)(3^2 + \gamma ^2)} -
\frac{\gamma ^2}{(1^2 + \gamma ^2)(2^2 + \gamma ^2)(3^2 + \gamma ^2)} \\ 
D_6 &=& \frac{11\gamma}{6(1^2 + \gamma ^2)(2^2 + \gamma ^2)(3^2 + \gamma
^2)} - \frac{\gamma ^3}{6(1^2 + \gamma ^2)(2^2 + \gamma ^2)(3^2 + \gamma
^2)} 
\end{eqnarray}
etc.,
then
\begin{equation}
Y_1(w) =A1
 \cos (\gamma \ln iw)  + A2 \sin (\gamma \ln iw)
\end{equation}
with
\begin{eqnarray}
A1&=&  1 + \frac{1}{1+\gamma ^2}\left(\frac{w}{2}\right)^2  +
\left\{ \frac{1}{(1^2 +\gamma ^2)(2^2 + \gamma ^2)} - \frac{\gamma
^2}{2(1^2 + \gamma ^2)(2^2 + \gamma ^2)}\right\}
\left(\frac{w}{2}\right)^4 \nonumber \\ &+& \left\{ \frac{1}{(1^2 + \gamma ^2)(2^2 +
\gamma ^2)(3^2 + \gamma ^2)} -\frac{\gamma ^2}{(1^2 +\gamma ^2)(2^2
+\gamma ^2)(3^2 +\gamma ^2)} \right\}\left(\frac{w}{2}\right)^6 + ....\nonumber \\
\end{eqnarray}
and
 \begin{eqnarray}
A2&=& \frac{\gamma}{1^2 +\gamma
^2}\left(\frac{w}{2}\right)^2 +\frac{3\gamma}{2(1^2 +\gamma ^2)(2^2
+\gamma ^2)}\left(\frac{w}{2}\right)^4 \nonumber \\ &+& \left\{-\frac{\gamma ^3}{6(1^2
+\gamma ^2)(2^2 +\gamma ^2)(3^2 +\gamma ^2)} + \frac{11\gamma}{6(1^2
+\gamma ^2)(2^2 +\gamma ^2)(3^2 +\gamma ^2)} 
\right\}\left(\frac{w}{2}\right)^6 + .... \nonumber \\
\end{eqnarray}
Now 
\begin{eqnarray}
\cos (\gamma \ln iw) = \cosh \left(\frac{\pi \gamma}{2}\right)
\cos(\gamma \ln w) - \sinh \left(\frac{\pi \gamma}{2}\right) \sin
(\gamma \ln w) \\
\sin (\gamma \ln iw) = \cosh \left(\frac{\pi \gamma}{2}\right)
\sin(\gamma \ln w) + \sinh \left(\frac{\pi \gamma}{2}\right) \cos
(\gamma \ln w)
\end{eqnarray}
Then
\begin{equation}
J_{i\nu}(iw)=Y_2(w)+iY_1(w)
\end{equation}
where $w$ and $\nu$ are real numbers.
Then the solution can be obtained from eqn.(104).
\section{Conclusion}
Since we have appended conclusions in each section, here
we would like to be very brief and only mention
what we have done in this article. Our overall study on the solutions of quantum mechanical equations in Rindler
space has been divided into five sections. In  section 2 we have made non-relativistic approximation and
solved exactly the Schr{\"{o}}dinger equation. We have predicted a kind of new quanta. We have named it as Cosmic
Phonon.

In section 3 with some kind of linear approximation, we have solved the Schr{\"{o}}dinger equation and shown that
the differential equation to which the Schr{\"{o}}dinger equation ultimately reduced is analogous to the
differential equation satisfied by the electrons emitted from the surface of a metal under the action of a strong
electric field. We have argued in that section that the coincidence is not accidental. The physical processes are
exactly identical. Although happening in an entirely different world.

In section 4 we made a quadratic approximation. We have solved the Schr{\"{o}}dinger equation and obtained eigen
values and the wave functions. We have noticed that the probability
densities are damped oscillatory in nature.
This typ of variation may be interpreted 
as the decrease in density of created 
particles as one goes away from the event horizon of a
black hole. Or in other words, the particle production decreases and finally vanishes as the distance from the event
horizon becomes large enough.

The exact relativistic form of the quantum mechanical equation has been studied in section 5. We have solved the
differential equation analytically and obtained the exact solution.
We have solved for the eigen functions and the energy eigen values as
a function of the free
parameter $\beta/m_0$. We have noticed that as the ratio approaches
$1$, system becomes more and more strongly bound- the energy
becomes highly negative and the range of the wave function decreases.

In section 5 we have studied the modified form of Kliein-Gordon
equation in Rindler space. We have solved the equation analytically
in an exact manner and obtained the eigen functions and the energy
eigen values. We have noticed that the
wave functions are given by Bessel functions with both imaginary orders and imaginary arguments. However, the energy
eigen values are found to be real.
\section{Appendix}
\noindent {\bf{A1:}} In this part of the Appendix we have used some of the
established conventional formulas of
special relativity with uniform accelerated motion and obtained
the single particle Lagrangian and
Hamiltonian in Rindler space.
Using the results from \cite{R5,R6,R7,R8} the Rindler coordinates are given by
\begin{eqnarray}
ct&=&\left (\frac{c^2}{\alpha}+x^\prime\right )\sinh\left (\frac{\alpha t^\prime}
{c}\right ) ~~{\rm{and}}~~ \nonumber \\
x&=&\left (\frac{c^2}{\alpha}+x^\prime\right )\cosh\left (\frac{\alpha t^\prime}
{c}\right ) 
\end{eqnarray}
Hence one can also express the inverse relations
\begin{equation}
ct^\prime=\frac{c^2}{2\alpha }\ln\left (\frac{x+ct}{x-ct}\right )
~~{\rm{and}}~~ x^\prime=(x^2-(ct)^2)^{1/2}-\frac{c^2}{\alpha }
\end{equation}
The Rindler space-time coordinates, given by eqns.(149) and (150) 
are then essentially an accelerated frame
transformation of the Minkowski metric of special relativity. The
Rindler coordinate transforms the Minkowski line element
\begin{eqnarray}
ds^2&=&d(ct)^2-dx^2-dy^2-dz^2 ~~{\rm{to}}~~\nonumber \\ ds^2&=&\left
(1+\frac{\alpha x^\prime}{c^2}\right)^2d(ct^\prime)^2-{dx^\prime}^2
-{dy^\prime}^2-{dz^\prime}^2
\end{eqnarray}
The general form of 
the metric tensor may then be written as
\begin{equation}
g^{\mu\nu}={\rm{diag}}\left (\left (1+\frac{\alpha x}{c^2}\right
)^2,-1,-1,-1\right )
\end{equation}
Now following the concept of relativistic dynamics of special theory
of relativity, the action
integral may be written as \cite{R1}
\begin{equation}
S=-\alpha_0 \int_a^b ds\equiv \int_a^b Ldt
\end{equation}
Then from eqn.(153) after putting $\alpha_0=-m_0 c$, where $m_0$ is the
rest mass of the particle, the Lagrangian of the particle is given by
\begin{equation}
L=-m_0c^2\left [\left ( 1+\frac{\alpha x}{c^2}\right )^2 -\frac{v^2}{c^2}
\right ]^{1/2}
\end{equation}
where $\vec v$ is the three velocity of the particle. The three momentum of the
particle is therefore given by
\begin{equation}
\vec p=\frac{\partial L}{\partial \vec v}, ~~ {\rm{or}}
\end{equation}
\begin{equation}
\vec p=\frac{m_0\vec v}{\left [ \left (1+\frac{\alpha x}{c^2} \right )^2
-\frac{v^2}{c^2} \right ]^{1/2}}
\end{equation}
Hence the Hamiltonian of the particle may be written as
\begin{equation}
H=\vec p.\vec v-L ~~ {\rm{or}}
\end{equation}
\begin{equation}
H=m_0c^2 \left (1+\frac{\alpha x}{c^2}\right ) \left (1+
\frac{p^2}{m_0^2c^2}\right )^{1/2}
\end{equation}
\noindent {\bf{A2:}} Consider the differential equation
\begin{equation}
\frac{d^2X}{d\xi^2}+\xi X=0
\end{equation}
To obtain an analytical solution, let us substitute $X(\xi)=\xi^n\psi(\xi)$, where $n$ is an unknown quantity. Then the above
differential equation reduces to
\begin{equation}
\xi^2 \frac{d^2 \psi}{d \xi^2} +2n \xi \frac{d\psi}{d\xi}+[n(n-1)+\xi^3]\psi=0
\end{equation}
Let $\xi=\beta z^{2/3}$, where $\beta$ is another unknown quantity. Then we have the reduced form of above
equation in the following form
\begin{equation}
z^2\frac{d^2 \psi}{dz^2}+\left ( n+\frac{1}{4}\right ) \frac{4}{3}z \frac{d \psi}{dz} +\frac{4}{9} [n(n-1)+
\beta^3z^2]\psi(z)=0
\end{equation}
Let us choose $n=1/2$, then we have
\begin{equation}
z^2\frac{d^2\psi}{dz^2}+z\frac{d\psi}{dz}+\left [ \frac{4}{9} \beta^3 z^2- \frac{1}{9}\right ]\psi(z)=0
\end{equation}
Finally choosing $\beta=(9/4)^{1/3}$, we get
\begin{equation}
z^2\frac{d^2\psi}{dz^2}+z\frac{d\psi}{dz} +\left ( z^2 -\frac{1}{9} \right )\psi(z)=0
\end{equation}
Comparing this differential equation with the standard form of Bessel 
equation
\begin{equation}
z^2\frac{d^2\psi}{dz^2}+z\frac{d\psi}{dz} +\left ( z^2 - \nu^2\right )\psi(z)=0
\end{equation}
whose solution is $J_\nu(z)$, Bessel function of order $\nu$
or $H_\nu^{(2)}(z)$, the second kind Hankel function of order $\nu$.
Then depending on the physical situation, 
we have the  appropriate solution for eqn.(163) as either
\begin{equation}
\psi(z)=J_{1/3}(z) ~~{\rm{or}}~~ \psi(z)=H_{1/3}^{(2)}(z)
\end{equation}
\noindent {\bf{A3:}} For $\lambda=0$, the reduced of the differential equation
given in eqn.(61) is given by
\begin{equation}
\frac{d^2X}{d\rho^2}+\frac{\rho^2}{4}X=0
\end{equation}
To have an analytical solution, we put $u=\rho/2^{1/2}$
Then the abpve differential equation reduces to
\begin{equation}
\frac{d^2X}{du^2}+u^2 X=0
\end{equation}
Let $X(u)=u^n\psi(u)$, where $n$ is an unknown parameter. On
substituting $X(u)$ in the above differential equation we get
\begin{equation}
u^2\frac{d^2\psi}{du^2}+2nu\frac{d\psi}{du}+ [n(n-1)+u^4] \psi=0
\end{equation}
We next put $u=\beta v^{1/2}$, where $\beta$ is another unknown
quantity and $v$ is the new variable. Rearranging the above
differential equation in terms of the new variable $v$, we have
\begin{equation}
v^2\frac{d^2\psi}{dv^2}+v\left ( n+\frac{1}{2}\right )
\frac{d\psi}{dv} +\frac{1}{4}[n(n-1)+\beta^4v^2]\psi =0
\end{equation}
To reduce this equation to an well known form of differential
equation satisfied by special function, we put $n=1/2$ and
$\beta=2^{1/2}$. Then we have the final form of the above
differential equation
\begin{equation}
v^2\frac{d^2\psi}{dv^2}+v\frac{d\psi}{dv} +\left ( v^2
-\frac{1}{16}\right )\psi=0
\end{equation}
The solution of this equation is $J_{1/4}(v)$. Hence
\begin{equation}
X(\rho)=\frac{\rho^{1/2}}{2^{1/4}}J_{1/4}\left
(\frac{\rho^2}{4}\right )
\end{equation}
\noindent {\bf{A4:}} We know \cite{R11}
\begin{eqnarray}
&&F\left (a , \frac{1}{2} ; a + \frac{1}{2} ; -x\right ) = \Gamma
\left (a+\frac{1}{2}\right )
\frac{x^{\frac{1-2a}{4}}}
{(1+ x )^{\frac{1}{2}}}P_{-\frac{1}{2}}^{\frac{1}{2} - a}
\left(\frac{1-x}{1+x}\right) \\
&&{\rm{Then}}~ G_{k-1}(m_0,\beta) =
\frac{\left[\Gamma(k+1)\right]^2}{(2m_0)^{\frac{1}{2}}(m_0 ^2 - \beta
^2)^{\frac{2k+1}{4}}} \nonumber \\ && \left[P_{-\frac{1}{2}}^{-\frac{1}{2}- k}
\left( \frac{\beta}{m_0}\right) + (-1)^k
P_{-\frac{1}{2}}^{-\frac{1}{2}-k} \left(-\frac{\beta}{m_0}\right) \right]
 \\
&& G_{k+2}(m_0,\beta) =
\frac{\left[\Gamma(k+4)\right]^2}{(2m_0)^{\frac{1}{2}}(m_0 ^2 - \beta
^2)^{\frac{2k+7}{4}}}\nonumber \\ && \left[P_{-\frac{1}{2}}^{-\frac{1}{2}- k -4}
\left( \frac{\beta}{m_0}\right) + (-1)^{k+3}
P_{-\frac{1}{2}}^{-\frac{1}{2}-k -4} \left(-\frac{\beta}{m_0}\right) \right]
\\
&& G_{k}(m_0,\beta) =
\frac{\left[\Gamma(k+2)\right]^2}{(2m_0)^{\frac{1}{2}}(m_0 ^2 - \beta
^2)^{\frac{2k+3}{4}}}\nonumber \\ && \left[P_{-\frac{1}{2}}^{-\frac{1}{2}- k -2}
\left( \frac{\beta}{m_0}\right) + (-1)^{k+1}
P_{-\frac{1}{2}}^{-\frac{1}{2}-k -2} \left(-\frac{\beta}{m_0}\right) \right]
\\ {\rm{ and}} \nonumber \\
&& G_{k+1}(m_0,\beta) =
\frac{\left[\Gamma(k+3)\right]^2}{(2m_0)^{\frac{1}{2}}(m_0 ^2 - \beta
^2)^{\frac{2k+5}{4}}} \nonumber \\ && \left[P_{-\frac{1}{2}}^{-\frac{1}{2}- k -3}
\left( \frac{\beta}{m_0}\right) + (-1)^{k+2}
P_{-\frac{1}{2}}^{-\frac{1}{2}-k -3} \left(-\frac{\beta}{m_0}\right) \right]
\\
\end{eqnarray}
Now
\begin{equation}
P_{\nu}^{\mu +2 }(x) = -2(\mu + 1) \frac{x}{( 1 - x^2)^{\frac{1}{2}}}
P_{\nu}^{\mu +1 }(x) + (\mu - \nu)(\mu + \nu + 1) p_{\nu}^{\mu}(x) 
\end{equation}
where $ \mu = \frac{1}{2} - k -4 , \nu = \frac{1}{2} ,x= \pm
\frac{\beta}{m_0} ~~and~~ \frac{x}{(1 - x^2 )^{\frac{1}{2}}} =
\frac{\beta}{(m_0 ^2 - \beta ^2)^{\frac{1}{2}}} $ \\
Then
\begin{eqnarray}
&& P_{-\frac{1}{2}}^{\frac{1}{2}-k-2}\left (\frac{\beta}{m_0}\right ) =
\frac{(2k+5)\beta}{(m_0 ^2 -\beta ^2)^{\frac{1}{2}}} + (k+3)^2
P_{-\frac{1}{2}}^{\frac{1}{2}-k-4}\left (\frac{\beta}{m_0}\right ) \\
&&{\rm{and}}\nonumber \\
&& G_{k+2}(m_0,\beta) = \frac{1}{(m_0 ^2 - \beta ^2)} \left[(k+2)^2
G_{k}(m_0,\beta)-(2k+5)\beta G_{k+1}(m_0,\beta) \right]  \nonumber \\
&& {\rm{Hence}}  \\
&& G_{k}(m_0,\beta) = \frac{1}{(m_0 ^2 - \beta ^2)} \left[k^2
G_{k-2}(m_0,\beta)-(2k+1)\beta G_{k-1}(m_0,\beta) \right] 
\end{eqnarray}
These are the recursion relations for the polynomials
$G_i(m_0,\beta)$ for various $i$.
To evaluate $G_{k}(m_0,\beta)$ for any arbitrary $k$ from the above
recursion relation, we use
\begin{eqnarray}
&& G_{0}(m_0 , \beta) = \frac{1}{(2m_0)^{1/2}(m_0 ^2 - \beta
^2)^{3/4}}P_{-1/2}^{-3/2}\left (\frac{\beta}{m_0}\right ) \\
&& G_{1}(m_0 , \beta) = \frac{1}{(2m_0)^{1/2}(m_0 ^2 - \beta
^2)^{7/4}} P_{-1/2}^{-5/2}\left (\frac{\beta}{m_0}\right ) 
\end{eqnarray}
Then
\begin{eqnarray}
&& X(-d_x ^2 + m_0 ^2)^{\frac{1}{2}} X^n \exp(-\beta X) = M_{n+1}(m_0 ,
\beta , X)\exp(-\beta X) \nonumber \\ &=& \left[ \sum_{k=0}^{n+1}F_{k,n+1}(m_0 , \beta)
X^{n+1-k} \right] \exp(-\beta X)
\end{eqnarray}
where
\begin{equation}
F_{k,n+1}(m_0,\beta) = \frac{1}{\pi ^ \frac{1}{2}}
\left(\begin{array}{c} n+1 \\k \end{array} \right) \left[ k
G_{k+2}(m_0,\beta)-\beta G_{k-1}(m_0,\beta) \right] \\
\end{equation}
Hence we have
\begin{eqnarray}
&& F_{k,n+2}(m_0,\beta) = \frac{1}{\pi^{1/2}}
\frac{(n+2)!}{k! (n+2-k)!} \nonumber \\ && \left[ k
G_{k+2}(m_0,\beta)-\beta G_{k-1}(m_0,\beta) \right] = \frac{n+2}{n+2-k}
F _{k,n+1} (m_0 ,\beta)\nonumber \\
\end{eqnarray}
We also have
\begin{eqnarray}
&& F_{0,n+1}(m_0 , \beta) = \frac{1}{(2 \pi m_0)^{1/2}(m_0 ^2 - \beta
^2)^{1/4}}P_{-\frac{1}{2}}^{-\frac{3}{2}}\left
(\frac{\beta}{m_0}\right ) \\
&& F_{1,n+1}(m_0,\beta)=\frac{-(n+1)\beta}{(2\pi m_0)^{1/2}(m_0 ^2 - \beta
^2)^{3/4}} P_{-\frac{1}{2}}^{-\frac{3}{2}}\left
(\frac{\beta}{m_0}\right ) \nonumber \\
\end{eqnarray}
Then for the quantum mechanical equation
\begin{equation}
X(-d_x ^2 + m_0 ^2)^{\frac{1}{2}} X^n \exp(-\beta X) = E \Psi(x) ,
\end{equation}
with
\begin{equation}
\Psi(X) \propto \sum_{k=1}^{n+1} \gamma_{k,n+1} X^n \exp(-\beta X) ,
\end{equation}
We have
\begin{equation}
X(-d_x ^2 + m_0 ^2)^{\frac{1}{2}} X^n \exp(-\beta X)= \left[
\sum_{k=0}^{n+1}F_{k,n+1}(m_0 , \beta) X^{n+1-k} \right] \exp(-\beta X)
\end{equation}
Then
\begin{eqnarray}
&& X(-d_x^2+m_0 ^2)^{\frac{1}{2}}\sum_{k=1}^{n+1}\gamma_{k,n+1}
X^k\exp(-\beta X) \nonumber \\
&=& \sum_{k=1}^{n+1} \gamma_{k,n+1}
\sum_{p=0}^{k}F_{p,k}(m_0 , \beta) X^{k+1-p} \exp(-\beta X) \nonumber \\
&=& E\sum_{k=1}^{n+1}\gamma_{k,n+1}X^k \exp(-\beta X) \nonumber \\
\end{eqnarray}
Hence equating the coefficients of $X^l$ from both the sides, after
putting $p=1$, we have
\begin{eqnarray}
&&\gamma_{l,n+1}(\beta,m_0) F_{1,l}(m_0 , \beta) = E_l(\beta,m_0) \gamma_{l,n+1}(\beta,m_0) \\
&& E_l (\beta,m_0)= F_{1,l} (m_0 ,\beta)
\end{eqnarray}
which gives the energy spectrum, provided the parameter $\beta$ is known.
Where
\begin{equation}
F_{1,l}(m_0,\beta)=\frac{-l\beta}{(2\pi m_0)^{1/2}(m_0 ^2 - \beta
^2)^{3/4}} P_{-\frac{1}{2}}^{-\frac{3}{2}}\left
(\frac{\beta}{m_0}\right ) \\
\end{equation}
with $l=1,2,3,......$, positive integers.
Therefore to obtain the energy spectrum, we have to evaluate the
associated Legendre function. It is to be noted further that the energy
spectrum is real and linearly quantized and the wave functions are
bounded ($\propto \exp(-beta X)$). From the
expressions it is quite obvious that $\beta < m_0$. This is also a
necessary condition for the argument $z$ of $P_\nu^\mu(z)$.
Now \cite{R11}
\begin{eqnarray}
&& P_{\nu}^{\mu}(z) = \frac{1}{\Gamma(1-\mu)} \left(\frac{z+1}{z-1}
\right)^{\frac{\mu}{2}} F\left (-\nu , \nu+1 ; 1-\mu ;
\frac{1-z}{2}\right ) \\
&&{\rm{with}} ~~ |1-z|< 2 
~~{\rm{where}}~ \cite{R11}~~\nonumber \\
&& F(a,b;c;d) = \frac{\Gamma(c)}{\Gamma(b)\Gamma(c-b)} \int_{0}^{1} t^{b-1}
(1-t)^{c-b-1} (1-tz)^{-a} dt 
\end{eqnarray}
with ${\rm{Re(c)}} > {\rm{Re(b)}} > 0$.
Here $ \mu = -3/2 , \nu = -1/2 ~{\rm{and}}~ z = \beta/ m_0 . $ Therefore 
\begin{equation}
P_{-1/2}^{-3/2} = \frac{1}{\Gamma(5/2)} \left( \frac{m_0+\beta}{m_0-\beta}
\right)^{-3/2} F\left(\frac{1}{2},\frac{1}{2}; \frac{5}{2};
\frac{m_0-\beta}{2m_0}\right)
\end{equation}
Now from eqn.(197), we have
\begin{equation}
F\left (\frac{1}{2},\frac{1}{2};\frac{5}{2};\frac{m_0 -
\beta}{2m_0}\right ) =
\frac{\Gamma(\frac{5}{2})}{\Gamma(\frac{1}{2})\Gamma(2)} \int_{0}^{1}
t^{-\frac{1}{2}}
(1-t)^{1}\left[1-t\left( \frac{m_0- \beta}{2m_0}\right)\right]^{-1/2}
dt \nonumber \\
\end{equation}
which may be decomposed into two integers, given by
\begin{equation}
F\left (\frac{1}{2},\frac{1}{2};\frac{5}{2};\frac{m_0 -
\beta}{2m_0}\right ) =
\frac{3(2m_0)^{1/2}}{4}(I_1 - I_2)
\end{equation}
where
\begin{equation}
I_1 = \int_{0}^{1} \frac{dt}{[at^2 +bt]^{1/2}}
\end{equation}
with $a = m_0-\beta  ~~{\rm{and}}~~ b=2m_0 . $ 
and
\begin{equation}
I_2 = \int_{0}^{1} t\frac{dt}{[at^2 +bt]^{1/2}}
\end{equation}
with same $a$ and $b$. The integrals
$I_1$ and $I_2$ can be evaluated analytically \cite{R12} and are given by
\begin{eqnarray}
&&I_1 = \frac{2(2m_0)^{1/2}}{\beta - m_0} \ln \left[ \frac{\beta -m_0
+\{(m_0-\beta)^2 +4m_0 ^2\}^{1/2}} {2m_0}\right] \\
&& {\rm{and}} \nonumber \\
&& I_2 = \frac{3(2m_0)^{5/2}}{(\beta - m_0)^{3}} \ln \left[ \frac{(\beta
-m_0)
+\{(m_0-\beta)^2 +4m_0 ^2\}^{1/2}}{2m_0} \right]\nonumber \\
&+&\frac{(2m_0)^{1/2}}{(\beta - m_0)^2} \left[(\beta - m_0)^2
+4m_0^2\right]^{1/2}  \nonumber \\
\end{eqnarray}
Then we have
\begin{eqnarray}
&& P_{-1/2}^{-3/2}\left (\frac{\beta}{m_0}\right )=\frac{(2m_0)}{\pi^{1/2} 
(m_0^2-\beta^2)^{3/4}}\nonumber \\ &&\left[(2\beta^2 -4\beta m - 10\beta ^2) \ln(P1)
-(m_0-\beta)(\beta ^2
- 2\beta m_0 + 5m_0 ^2 )^{1/2} \right] \nonumber \\
\end{eqnarray}
where 
\begin{equation}
P1=
\frac{m_0-\beta +\{( m_0-\beta)^2 +4m_0 ^2\}^{1/2}}{2m_0} 
\end{equation}
We also need $P_{-1/2}^{-5/2}(\beta /m_0)$, which is given by
\begin{eqnarray}
P_{-1/2}^{-5/2} = \frac{1}{\Gamma(7/2)} \left(
\frac{m_0+\beta}{m_0-\beta}
\right)^{-5/4} F\left(\frac{1}{2},\frac{1}{2}; \frac{7}{2};
\frac{m_0 -\beta}{2m_0}\right) \nonumber \\
\end{eqnarray}
where
\begin{equation}
F\left (\frac{1}{2},\frac{1}{2};\frac{7}{2};\frac{m_0 -
\beta}{2m_0}\right ) =
\frac{\Gamma(\frac{7}{2})}{\Gamma(\frac{1}{2})\Gamma(3)} \int_{0}^{1}
\frac{(1-t)^{2}}{\left[2m_0 t+ (m_0- \beta)t^2\right]^{1/2}} dt
\nonumber \\
\end{equation}
The integral may be decomposed into three parts and can easily be
evaluated analytically \cite{R12}
\begin{eqnarray}
I_1 &=& \int_{0}^{1} \frac{dt}{[at^2 +bt]^{1/2}}
= \frac{2(2m_0)^{1/2}}{m_0-\beta } \nonumber \\ && \ln \left[ \frac{m_0-\beta
+\{(m_0-\beta)^2 +4m_0 ^2\}^{1/2}} {2m_0}\right], 
\end{eqnarray}
\begin{eqnarray}
I_2 &=& -2 \int_{0}^{1} \frac{t dt}{[at^2 +bt]^{1/2}}
= -\frac{6(2m_0)^{5/2}}{(m_0-\beta)^{3}} \nonumber \\ && \ln \left[ \frac{(
-m_0-\beta)
+\{(m_0-\beta)^2 +4m_0 ^2\}^{1/2}}{2m_0} \right] \nonumber \\
&-&\frac{(8m_0)^{1/2}}{(m_0-\beta)^2} \left[(m_0-\beta )^2
+4m_0^2\right]^{1/2}  
\end{eqnarray}
and
\begin{equation}
I_3 = \int_{0}^{1} \frac{t^2 dt}{[at^2 +bt]^{1/2}}
= \frac{2(2m_0)^{9/2}}{(m_0-\beta)^{3}}  \ln(A1+A2+A3)
\end{equation}
where $a$ and $b$ are same as before and
\begin{equation}
A1=\frac{19}{8} \frac{(m_0-\beta)
+\{(m_0-\beta)^2 +4m_0 ^2\}^{1/2}}{2m_0},
\end{equation}
\begin{equation}
A2=\frac{11}{8}\frac{(m_0-\beta)}{2m_0} \left\{(m_0-\beta )^2
+4m_0^2\right\}^{1/2}
\end{equation}
and
\begin{equation}
A3=\frac{(m_0-\beta)}{4(2m_0)^{5/2}} \{(m_0-\beta)^2
+4m_0 ^2\}^{3/2} 
\end{equation}
Now to obtain the eigen states, we put $p=0$ in eqn.(192) and equate the
coefficients of $x^l$ from both the sides and obtain
\begin{equation}
\gamma_{l,n+1}(\beta,m_0)F_{0,l}(\beta, m_0) = E_{l+1}(\beta , m_0)\gamma_{l+1,n+1}(\beta,m_0)
\end{equation}
Since $F_{0,l}(\beta , m_0)$ is independent of $l$, we put it as
$F(\beta , m_0)$. Hence
\begin{equation}
\gamma_{l+1,n+1} (\beta,m_0)= \frac{F(\beta,m_0)}{E_{l+1}(\beta,m_0)}\gamma_{l,n+1}(\beta,m_0)
\end{equation}
Putting the value of $E_{l+1}(\beta , m_0)$, we get
\begin{equation}
\gamma_{l+1,n+1}(\beta,m_0) = \frac{F(\beta,m_0)}{F_{1,l+1}(\beta,m_0)}\gamma_{l,n+1}
\end{equation}
This is the recursion relation for $\gamma_{l,n+1}(\beta,m_0)$.
Hence we have
\begin{eqnarray}
\gamma_{2,n+1}(\beta,m_0) &=& \frac{F(\beta
,m_0)}{F_{1,2}(\beta,m_0)}\gamma_{1,n+1}(\beta , m_0) \\
\gamma_{3,n+1}(\beta,m_0) &=& \frac{F(\beta
,m_0)}{F_{1,3}(\beta,m_0)}\gamma_{2,n+1}(\beta , m_0)\nonumber \\  &=& \frac{F^2 (\beta
, m_0)}{F_{1,2}(\beta, m_0) F_{1,3}(\beta, m_0)}\gamma_{1,n+1}(\beta
,m_0)\\
\gamma_{4,n+1} (\beta,m_0)&=& \frac{F^3 (\beta
, m_0)}{F_{1,2}(\beta, m_0) F_{1,3}(\beta, m_0) F_{1,4}(\beta,m_0)}\gamma_{1,n+1}(\beta
,m_0)\\
\gamma_{l,n+1}(\beta,m_0) &=& \frac{F^{l-1} (\beta,m_0)} {\prod^l_{i=2}F_{i,4}(\beta,m_0)}\gamma_{1,n+1}(\beta
,m_0)
\end{eqnarray}
The last one is the most general expression.
Therefore the wave function can be expressed as
\begin{eqnarray}
\Psi(X) &=&\Psi_n (X)= \sum_{l=1}^{n+1} \gamma_{l,n+1} X^l \exp(-\beta X) \nonumber \\
&=&\left[\sum_{l=1}^{n+1}\frac{F^{l-1}(\beta,
m_0)}{\prod^l_{i=1}F_{1,i}(\beta, m_0)}
\right]\gamma_{1,n+1}(\beta,m_0) X^l \exp(-\beta X)
\end{eqnarray}
The normalization constant $\gamma_{1,n+1}(\beta,m_0)$ can be obtained
from the orthonormality condition
\begin{equation}
\int_{1}^{\infty} \Psi _n ^* (X) \Psi_{n^\prime} (X) dX =\delta_{nn^\prime}
\end{equation}
Which gives
\begin{eqnarray}
\gamma_{1,n+1}^{2}(\beta ,m_0) &=& \left[\sum_{l,l^\prime
=1}^{n+1}\frac{F^{l-1}(\beta , m_0)F^{l^\prime -1}(\beta ,
m_0)}{\prod^l_{i=1}F_{1,i}(\beta , m_0)
\prod^{l^\prime}_{j=1}F_{1,j}(\beta,m_0)}
\right] \times \nonumber \\ &&\left[\frac{(l+l^\prime)!-E_\gamma (\beta ,l ,l^\prime)
}{2^{l+l^\prime +1} \beta ^{l+l^\prime +1}} \right]
\end{eqnarray}
Hence the normalization  constant is the square root of this quantity.
Here
\begin{equation}
E_\gamma (\beta , l , l^\prime ) = \int_{0}^{2\beta}
z^{l+l^\prime}\exp(-z) dz
\end{equation}
the incomplete $\Gamma$-function.

\begin{figure}[ht]
\psfig{figure=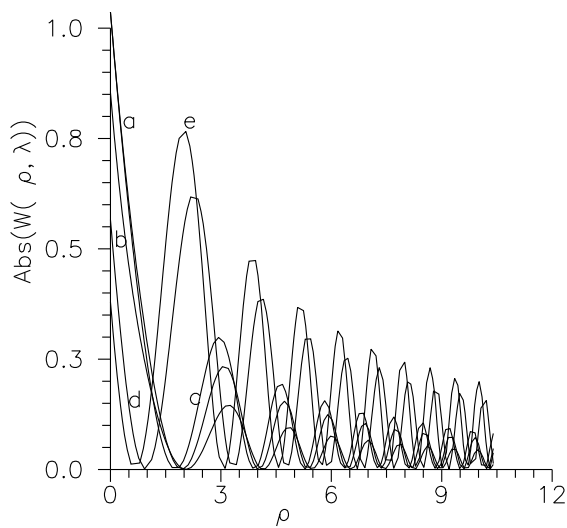,height=0.5\linewidth}
\caption{
The variation of $\mid W(\rho,\lambda)\mid^2$, 
the probability density with $\rho$, for $\lambda=0.5$, curve (a),
$0.25$, curve (b), $-0.05$, curve (c), $-0.5$, curve (d) and $-0.75$,
curve (e) in units of $q/4$. 
}
\end{figure}
\begin{figure}[ht]
\psfig{figure=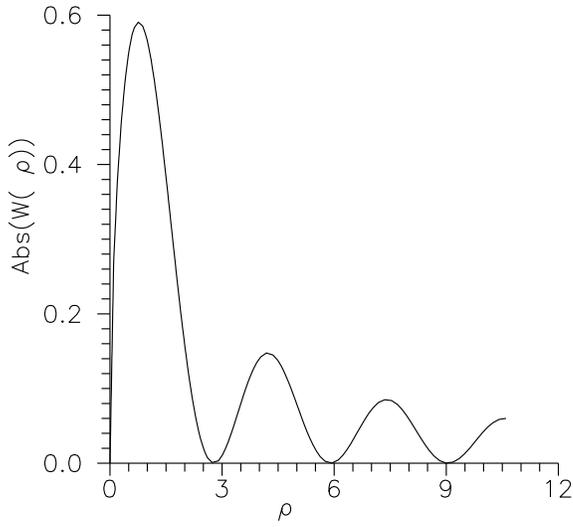,height=0.5\linewidth}
\caption{
The variation of
$\mid X(\rho)\mid^2$ with $\rho$.
}
\end{figure}
\begin{figure}[ht]
\psfig{figure=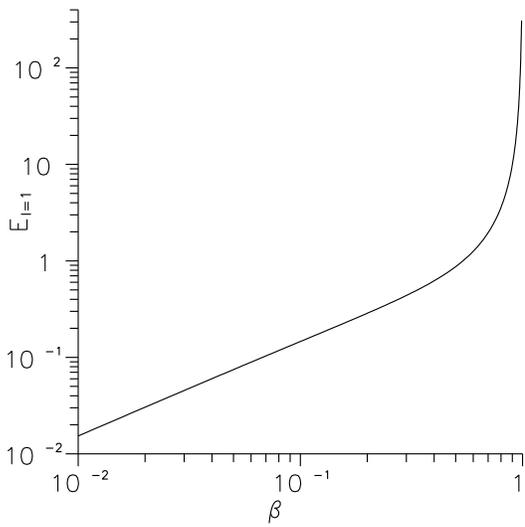,height=0.5\linewidth}
\caption{
The variation of $E_l$ with $\beta$ for $l=1$. Here the parameter $\beta$ is re-defined as $\beta/m_0$.
}
\end{figure}
\end{document}